# Quantum computation and hidden variables.[*]

V.V. Aristov and A.V. Nikulov[+]

Institute of Microelectronics Technology, Russian Academy of Sciences, 142432 Chernogolovka, Moscow District, Russia

**ABSTRACT**

Many physicists limit oneself to an instrumentalist description of quantum phenomena and ignore the problems of foundation and interpretation of quantum mechanics. This instrumentalist approach results to "specialization barbarism" and mass delusion concerning the problem, how a quantum computer can be made. The idea of quantum computation can be described within the limits of quantum formalism. But in order to understand how this idea can be put into practice one should realize the question: "What could the quantum formalism describe?", in spite of the absence of an universally recognized answer. Only a realization of this question and the undecided problem of quantum foundations allows to see in which quantum systems the superposition and EPR correlation could be expected. Because of the "specialization barbarism" many authors are sure that Bell proved full impossibility of any hidden-variables interpretation. Therefore it is important to emphasize that in reality Bell has restricted to validity limits of the no- hidden-variables proof and has shown that two-state quantum system can be described by hidden variables. The later means that no experimental result obtained on two-state quantum system can prove the existence of superposition and violation of the realism. One should not assume before unambiguous experimental evidence that any two-state quantum system is quantum bit. No experimental evidence of superposition of macroscopically distinct quantum states and of a quantum bit on base of superconductor structure was obtained for the present. Moreover same experimental results can not be described in the limits of the quantum formalism.

## 1. INTRODUCTION

The quantum computation is one of the most grandiose and inspiring ideas of our time. But it is very difficult not only to make quantum computer but even to understand how it can be made. On the one hand the idea of the quantum computation can be described in the limits of the quantum formalism, but on the other hand the history of its emergence is bound up with the battle over the interpretation of quantum mechanics. This battle raged until the early 1930s. The creators, Albert Einstein, Niels Bohr and others, well realized the magnitude of the questions of foundations of the quantum mechanics. But after the 1930s there followed a long period in which most physicists turned their attention elsewhere, and progress in understanding the foundations of quantum mechanics attracted only the attention of the relatively small number of people who continued to seek an understanding of these matters. During this period, the wonderful difficulties of quantum mechanics were largely trivialized, swept aside as unimportant philosophical distractions by the bulk of the physics community. The interest to the problem the foundations of quantum mechanics was renewed by a new understanding of the Einstein - Podolsky - Rosen paradox[1] inspired of the Bell's theorem[2] and its experimental tests[3-6]. But as, N.D. Mermin suggested in 1985[7], in the question of whether there is some fundamental problem with quantum mechanics signaled by tests of Bell's inequality, physicists can be divided into a majority who are "indifferent" and a minority who are "bothered". This division on the "indifferent majority" and the "bothered minority" can be observed now in two types of publications concerning the problem of the quantum computation.

One can see from some publications, which we will ascribe to the first type, that their authors do not understand enough profoundly the problems of foundations of quantum mechanics and limit oneself to the quantum formalism.

---

Nevertheless, these authors (for example of the book[8] and review[9]), which may be ascribed to the "indifferent majority", mention on the Bell's theorem and the EPR paradox because of their popularity. But in the first type of publications these problems is mentioned only in passing and their connection with the problem of the quantum computation is not explored enough profoundly. Moreover, the problem of foundations of quantum mechanics is understood incorrectly in some publications of the first type. For example, the authors of the book[8] reproach Einstein, Podolsky and Rosen with their aspiration to force the physical community to return to classical conception about laws of the Nature. Unfortunately not only these authors but also most physicists appertain to the "indifferent majority" do not understand the fundamental importance of the EPR paradox for our understanding of quantum mechanics and for the problem of quantum computation.

The "bothered minority" of physicists realize the wonderful difficulties of quantum mechanics, revealed in particular by the EPR paradox, and make efforts to clear its mystery in the publications (see for example[10-12]), which we will ascribe to the second type. But for the present these efforts rather obscure than clear the mystery because of difference of opinion about the foundations of quantum mechanics (see for example the debates at the conferences[13]). The experts must admit that *quantum mechanics is not yet based on a generally accepted conceptual foundation*[14].

Since the debates on the foundations of quantum mechanics never reached a satisfactory conclusion the disinclination of the authors[8,9], and many others, to wallow in vain debates[13] seems, on the face of it, quite wise. It seems to debate a philosophical matter where are only questions and no any satisfactory answers unreasonable at the consideration of a practical problem of quantum computer. But it is important in some cases to understand questions, even if any satisfactory answers are absent. The quantum computer is just such case since the idea of the quantum computation has emerged as a by-product of the long-term ineffectual philosophical debates on the foundations of quantum mechanics. Unfortunately these debates are insufficiently studied in school. This defect of teaching is main reason of the prevalence of the first type of attitude to the problem of quantum computations. Any physicist graduating university has learned the quantum formalism but he can not give a satisfactory answer on the question: "What could this formalism describe?" Moreover many physicists reject such question as useless philosophy. They, including the authors of the first type of publications[8,9], seem to think that not only the question on the object of description but even the difference between the quantum description and the object of description is philosophical distractions. Because of the enormous successfulness in describing of nature at the atomic (and not only atomic) level many physicists are sure that one should not come to nothing more than the scholastic (studied in a school) quantum formalism in order to understand how a quantum computer can be made. But such scholastic approach results to some delusions. The authors of the publications[8,9] and others describe enough well the idea of quantum computations using the quantum formalism. But in order to realize this idea one should deal with any objects described by the quantum formalism. That is in order to make quantum computer we should understand philosophical problems of quantum mechanics which were the point at issue of Einstein and Bohr.

The main point of the disagreement between Einstein and Bohr was the problem of completeness of quantum mechanics and a possibility of a more full theory of objective reality. Einstein emphasized that for centuries science had viewed its aim as the discovery of the *real*. In 1949, responding to the accolades lavished in honor of his seventieth birthday, he emphasized that the quantum theory in its universally recognized Copenhagen interpretation had relinquished precisely what has always been the goal of science: "*the complete description of any (individual) real situation (as it supposedly exists irrespective of any act of observation or substantiation)*."[15] In contrast to this Bohr stated that all hope of attaining a unified picture of objective reality must be abandoned. According to his point of view quantum theory would provide predictions concerning the results of measurements but, unlike all previous theories, it was incapable of providing a full account of "how nature did it." Bohr argued that the very desire to seek such a complete account was misguided and naive.

In fact, the debate between Einstein and Bohr was the debate about an existence of objective reality. In order to understand a deep philosophical essence of this debate and its importance for the problem of quantum computer we should realize a fundamental difference between the existence of objective reality and measurability of its parameters. It should be emphasized since, as some philosophers note, many physicists do not realize this difference. For example the celebrated polymath, Mortimer Adler, declares in[16] that "*Most theoretical physicists are guilty of failing to distinguish between a measurable indeterminacy and the epistemic indeterminability of what is in reality determinate. The indeterminacy discovered by physical measurements of subatomic phenomena simply tells us that we cannot know the definite position and velocity of an electron at one instant of time. It does not tell us that the electron, at any instant of time, does not have a definite position and velocity. Physicists convert what is not*

*measurable by them into the unreal and the nonexistent*". The difference between measurability and existence underlies the conception of hidden variables: such variable has a value preexisting a measurement but it is hidden because this value can not be revealed at the measurement. The problem of a possibility of hidden-variables interpretation of quantum mechanics has long and intricate history. And now it is misunderstood by many physicists. The main aim of the present paper is to consider this problem in connection with the problem of quantum computation. The main questions under consideration are: "Could a quantum system be used as quantum bit if it can be described in terms of hidden variables?" and "What experimental results can guarantee that the quantum system can not be described in terms of hidden variables?" In order to understand the problem to the best of our abilities we will consider in the next section the essence of the controversy on foundation of quantum mechanics.

## 2. ESSENCE OF THE CONTROVERSY ON FOUNDATION OF QUANTUM MECHANICS

Roger Penrose remarked[17] that while the quantum theory agrees incredibly well with experiment and while it is of profound mathematical beauty, it "makes absolutely no sense". These remark made by the famous physicist in 1986 defines exactly and correctly both the modern situation and the whole history of quantum mechanics. The quantum postulates were emerged in order to describe the paradoxical situation with which physicists are faced in atomic world. First quantum principles, the Plank's quantization proposed in 1900 and the Bohr's quantization proposed in 1913 may be considered as purely empirical. The modern quantum mechanics emerged in the remarkable period of 1925-1927 from the work of Heisenberg, de Broglie, Schrodinger and Born. The mathematical formalism of quantum mechanics, although refined and generalized in the intervening decades, has never been seriously challenged either theoretically or experimentally and remains as firmly established for the present as it was right from its beginning. Yet over the entire period since original development of quantum formalism there has been controversy about its interpretation. This controversy results from the refusal of quantum formalism to address certain issues.

### 2.1 Double-slit interference experiments

For example the quantum theory never describes how a particle can make its way through two slits at the same time in the interference experiment. Experiments demonstrate that electrons[18], neutrons[19], atoms[20], and even large molecules[21] passing one after another through two slits give the interference pattern on a detector screen in complete agreement with quantum formalism. But the quantum formalism refuses to answer on the question: "How can a single particle, which we observed in the source and in the detecting screen as being well-localized and much smaller than width of a single slit in the barrier, acquire information about the other very remote slit, if it was considered to pass only one through the slits?" Richard Feynman emphasized that the double-slit interference experiment is at the heart of quantum mechanics[22]: "In reality, it contains the only mystery, the basic peculiarities of all of quantum mechanics". Indeed, this experiment demonstrates very clear both advances and defects of universally recognized quantum formalism.

### 2.2 Duality

There is unambiguous experimental evidence of a double nature of a particle, at least an enough small particle. Whereas in classical physics there is a unity, in quantum mechanics this unity is replaced by a duality. Bohr, the first to identify this curious feature, termed it *complementarity*.

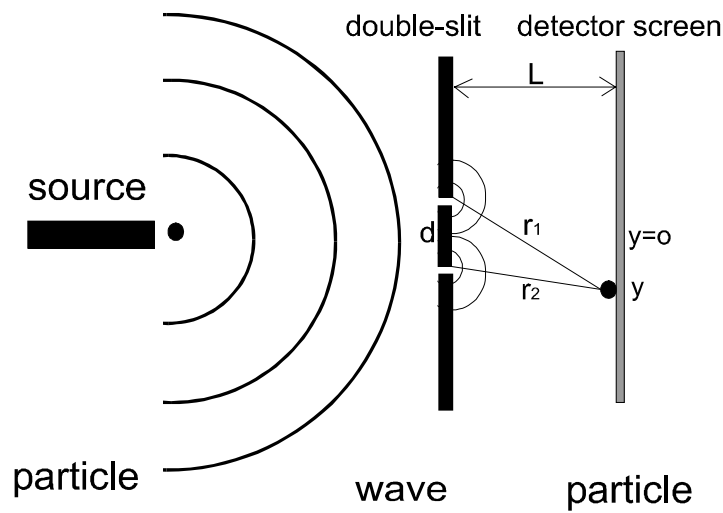

Fig. 1. Experimental evidence of duality in the double-slit interference experiment. A particle reveals itself as a localized object in a source and at a detecting screen. But in order to explain the interference pattern of the probability distribution of particle arrival on different places of the detecting screen we should conclude that the particle is also a wave in a some sense.

Experiment shows that each particle makes a single click in a well-localized point of a detecting screen. We can make sure that single particle flying out a source makes a single click in a well-localized point independently of distance between the source and the detecting screen. Thus, the experiment demonstrates that a particle is a well-localized object when it flies out the source and strikes to the detecting screen, Fig.1. But experiments[18-21] show also that the clicks of many particle passing through double slits are distributed along the detecting screen in accordance with the interference pattern predicted by a wave theory. Therefore we should conclude that the particle is also a wave in a some sense. Moreover the experiments[18-21] corroborate the conjecture of a genius de Broglie on matter wave with a wavelength

$$\lambda = \frac{2\pi\hbar}{p} \qquad (1)$$

connected with momentum $p = mv$ of the particle and the Born's interpretation of this wave as an amplitude of probability $P(r) = |\Psi(r)|^2$. According to the wave mechanics each of the two slits is a source of waves $\Psi_1(r) = A_1 exp(i\varphi_1)$ and $\Psi_2(r) = A_2 exp(i\varphi_2)$, Fig.1. According to the Born's interpretation that a particle will strike in a point $y$ of the detecting screen, Fig.1, with a probability

$$P(y) = |\Psi_1 + \Psi_2|^2 = A_1^2 + A_2^2 + 2A_1 A_2 \cos(\varphi_1 - \varphi_2) \qquad (2)$$

where $\varphi_1 = 2\pi r_1/\lambda = 2\pi(L^2+(y-d/2)^2)^{1/2}/\lambda$, $\varphi_2 = 2\pi r_2/\lambda = 2\pi(L^2+(y-d/2)^2)^{1/2}/\lambda$ and the phase difference $\varphi_1 - \varphi_2 \approx 2\pi dy/\lambda L$ when the distance $d$ between slits is much smaller than the distanced $L$ between the screen with double-slits and the detecting screen, Fig.1. The experiments corroborate this periodical distribution (2) $P(y) \propto 1 - cos(2\pi dy/\lambda L)$ of hits to the detecting screen of electrons[18], neutrons[19], atoms[20], and even large molecules[21], with period $\lambda L/d$ agreeing to the de Broglie's relation (1) for the wavelength.

2.3 Non-locality and collapse of the wave function

But this agreement between the quantum description and the experimental results can not eliminate the mystery of the double-slit interference. The quantum formalism refuses to answer on the question: "How and when could the wave turn into the particle?" Moreover it refuses to answer: "What could the wave function describe?" The wave is non-local and the particle is local. And the double-slit interference experiment demonstrates a transformation from the non-local object into the local one or a non-locality. Locality has long time a guiding principle of physics. This principle assumes that things done at one location only have effects at that location. But the double-slit interference experiment demonstrates obvious non-locality if we refuse to consider this experiment as a result of a transformation of a particle into a wave and backwards. If we drop the locality principle, and allow for the possibility that plugging one slit might alter the paths of particles passing through the other slit, we would no longer be forced to conclude this transformation or that particles can be in two places at once. Recent years have seen a growing interest in the consequences of dropping the locality principle.

Von Neumann described the transformation from the non-local wave into the local particle with help of a projection postulate[23]. This postulate is an inalienable part of quantum formalism independently of interpretation. The description of a quantum state with help of the wave function or the state vector can not be valid without this postulate. According to the quantum formalism a quantum system can be in a superposition of different states. For example, the wave function $\Psi(r)$ in the double-slit interference experiment describes the superposition of different location $r$ of the particle or the state vector

$$\psi = \alpha\psi_\uparrow + \beta\psi_\downarrow \qquad (3)$$

describes the superposition of eigenstates with different projection along a direction of a single spin-1/2. We can not see a particle simultaneously in different points or different value of spin projection. Therefore a collapse of the wave function $\Psi(r)$ or a reduction of the state vector (3) must be postulated at measurement. This "jump", just jump, from the wave function to the observed values, i.e. from possibility to reality, is not described in any way by the quantum formalism. This main unsettled problem results in different paradoxes and provokes different interpretations of the quantum mystery.

2.4 Non-locality and collapse of the wave function in the EPR paradox

The EPR paradox[1] emerges inevitably when a conservation law is added to the collapse of the wave function $\Psi(r)$ describing a quantum system, for example of two particle $\Psi(r_1,r_2) = \Sigma_p a_p exp[i(pr_1-pr_2)/\hbar]$, having zero common momentum $p_1 = -p_2 = p$. The collapse of this wave function at the measurement, for example, of the momentum of the first particle $p_1 = p\pm\Delta p$ with a uncertainty $\Delta p$, means a change of the state not only of this particle but also of the second particle $p_2 = -(p \pm \Delta p)$. This change should occur independently of the distance $r_1 - r_2$ between the particles or the time $(r_1 - r_2)/c$ because the conservation law can not depend on the distance or the time. Thus, the collapse of the wave function results to a "spooky action at a distance."

Einstein never accepted such "spooky action at a distance" and can not considered quantum mechanics as a complete theory. The EPR paradox[1] should demonstrate this incompleteness of quantum mechanics, basing on the two assumption, realism and locality, seemed obvious. Most physicists appertained to the "indifferent majority" are sure that this Einstein's attack on the Copenhagen interpretation defend by Bohr was a complete failure[8] and they regard the EPR paradox as an illustration of how quantum mechanics violates classical intuitions. But we think that the "indifferent majority" do not understand enough clear what means violation of locality or absence of reality and therefore are ready to see the quantum superposition in any quantum system even without any experimental evidence.

John Bell had the contrary position on this Bohr-Einstein debate[1,24]. He wrote: "I felt that Einstein's intellectual superiority over Bohr, in this instance, was enormous; a vast gulf between the man who saw clearly what was needed, and the obscurantist." The position of some other experts was also contrary to the one of the "indifferent majority". The first, who appreciated the EPR paradox at its true value was Schrodinger.

2.5 Collapse of the wave function and the Schrodinger's cat

Motivated by the EPR work[1] Erwin Schrodinger in the paper[25] entitled "The present situation in quantum mechanics" wrote: "Maximal knowledge of a total system does not necessarily include total knowledge of all its parts, not even when these are fully separated from each other and at the moment are not influencing each other at all" and he coined the term "entanglement of our knowledge" to describe this situation. Schrodinger used in the German original[25] "Verschrankung" and he himself introduced the English translation "entanglement" in the paper[26]. The entanglement, according to Schrodinger the essence of quantum mechanics, is at the heart of the EPR paradox. Although the entanglement appears directly from the quantum formalism only the EPR work[1] has revealed its paradoxicality. Therefore this paradoxical quantum phenomenon is called also EPR correlation. In order to illustrate this, let us consider a simplified version of the EPR thought experiment put forth by David Bohm[27]. The states of two particles with spin-1/2 are entangled in accordance with the Schrodinger's definition when, for example, we know that their total spin equals zero but we can not know the spin projection of each particle along any direction. The wave function may be written in this case as

$$\psi = \alpha \psi_\uparrow(r_1) \psi_\downarrow(r_2) + \beta \psi_\downarrow(r_1) \psi_\uparrow(r_2) \qquad (4)$$

According to the von Neumann postulate on the reduction of the state vector (4) the measurement of the spin projection of one particle along any direction results to the state of the other particle independently of the distance $r_1 - r_2$ between the particles: before the measurement of the second particle the probability of $\uparrow$- state of the first particle was $|\alpha|^2$ and after it has become 1 (at $\downarrow$- result for the second particle) or 0 (at $\uparrow$- result for the second particle). There is no paradox if the state vector (4) is interpreted as a description of our knowledge. But the paradox is obvious at any realistic interpretation since the spin projection can be measured along any direction.

Schrodinger proposed in the paper[26] other paradox connected with the entanglement and well known as Schrodinger's cat. This paradox is discussed already seventy years but for the present physicists did not achieve any universally recognized interpretation of it. It is important to reproduce punctually the remarkable paragraph describing the paradox in order to understand the Schrodinger's thought

*"One can even set up quite ridiculous cases. A cat is penned up in a steel chamber, along with the following diabolical device (which must be secured against direct interference by the cat): in a Geiger counter there is a tiny bit of radioactive substance, so small that perhaps in the course of one hour one of the atoms decays, but also, with equal probability, perhaps none; if it happens, the [Geiger] counter tube discharges and through a relay releases a hammer which shatters a small flask of hydrocyanic acid. If one has left this entire system to itself for an hour, one would say that the cat still lives if meanwhile no atom has decayed. The first atomic decay would have poisoned it. The ψ-function of the entire system would express this by having in it the living and the dead cat (pardon the expression) mixed or smeared out in equal parts."*

We can easy describe this paradoxical situation in terms of the quantum formalism by the state vector

$$\psi = \alpha \, at_{no} \, Ge_{no} \, fl_{no} \, cat_{live} + \beta \, at_{yes} \, Ge_{yes} \, fl_{yes} \, cat_{dead} \qquad (5)$$

of the entire system consisting of atom (the state of which is described by the state vector *at*), a Geiger counter (the state of which is described by the state vector *Ge*), a small flask (the state of which is described by the state vector *fl*) and a cat (the state of which is described by the state vector *cat*). The states of these system components are entangled both in the Schrodinger's description and in the state vector (5). Schrodinger assumed that the Geiger counter tube should without fail discharge $Ge_{yes}$ when one of the atoms has decayed $at_{yes}$ and it should not discharge $Ge_{no}$ if no atom has decayed $at_{no}$. Therefore products $at_{yes} Ge_{no}$ or $at_{no} Ge_{yes}$ are absent in the superposition (5). The same strong causality is assumed between the states of the Geiger counter, *Ge*, and the flask, *fl*, and between the states of the flask, *fl*, and the cat, *cat*. Therefore products $Ge_{yas} fl_{no}$, $Ge_{no} fl_{yes}$, $fl_{yes} cat_{live}$ and $fl_{no} cat_{dead}$ are absent also in the superposition (5). The same strongly determinate correlation between spin states of the two particle in the Bohm version[27] of the EPR paradox exists because of the law conservation. Therefore products $\psi_\uparrow(r_1)\psi_\uparrow(r_2)$ and $\psi_\downarrow(r_1)\psi_\downarrow(r_2)$ are absent in the superposition (4).

Thus, the entanglement results from a causality or other correlation between different parts of quantum system. Causality denoting the relationship between one event (called cause) and another event (called effect) is base of scientific knowledge and the correlation between states is not paradox. The EPR and Schrodinger's cat paradoxes result from the quantum superposition and its reduction at measurement assumed in quantum mechanics. The coherent superposition can explain the double-slit interference experiment and many other quantum phenomena. But its instantaneous reduction is very strange, results in many paradoxes and is not described in any way in the framework of quantum formalism. Therefore Einstein was right when he stated that the quantum mechanics is not complete theory. The experimental evidence[3-6] of violation of the Bell's inequalities can prove violation of the local realism. But it is not correct to think (as the "indifferent majority" is prone to do) that the experiments can prove the completeness of the quantum theory. The quantum formalism describes a something with help of wave functions and state vectors and can predict the statistical expectation of a observable. But the inexplicable "jump" from the state vector to the observable do not describe in any way. There is unclear also what is the something that the quantum formalism describes. Therefore quantum mechanics was and remains a mystery. This mystery has provoked numerous interpretations and warrants the search of new interpretations.

2.6 Different interpretations of the quantum mystery

The interpretations of various nature were proposed during 80 years[28]. The Copenhagen interpretation was formulated in 1927 by Bohr, Heisenberg and other[29]. David Bohm proposed in 1952[30] an interpretation in which the existence of a non-local universal wave function allows distant particles to interact instantaneously. This Bohm's interpretation generalizes Louis de Broglie's pilot wave theory from 1927, which posits that both wave and particle are real. In five years, in 1957, Hugh Everett[31] proposed a interpretation which was popularized later by B. S. DeWitt[32] as "many worlds interpretation". Many other versions of interpretations were proposed also, for example: the statistical interpretation[33], the transactional interpretation[34], the information-theoretic interpretation[35], the Vaxjo interpretation[36] and the anti-Vaxjo interpretation[37].

The Copenhagen interpretation is widely accepted amongst physicists. According to it, the probabilistic nature of quantum mechanics predictions cannot be explained in terms of some other deterministic theory, and does not simply reflect our limited knowledge. Quantum mechanics provides probabilistic results because the physical universe is itself probabilistic rather than deterministic. In contrast to the Copenhagen interpretation, the Bohm's interpretation[30] of quantum mechanics, sometimes called Bohmian mechanics, is the ontological and causal interpretation using hidden variables. In fact, the Bohmian interpretation opts for keeping realism and accepting a real non-locality. Within Bohm's interpretation, it can occur that events happening at one location in space instantaneously influence other events which might be at large distances: thus we say that the theory fails to obey locality, i.e., it is non-local. The response many physicists have to Bohm's theory is often related to how they regard this concept.

The Everett many-worlds interpretation holds that all the possibilities described by quantum theory simultaneously occur in a "multiverse" composed of mostly independent parallel universes. This is not accomplished by introducing some new axiom to quantum mechanics, but on the contrary by removing the axiom of the collapse of the wave packet: All the possible consistent states of the measured system and the measuring apparatus (including the observer) are present in a real physical (not just formally mathematical, as in other interpretations) quantum superposition. (Such a superposition of consistent state combinations of different systems is called an entangled state.) While the multiverse is deterministic, we perceive non-deterministic behavior governed by probabilities, because we can observe only the universe, i.e. the consistent state contribution to the mentioned superposition, we inhabit.

The real non-locality in the Bohm's interpretation and the Everett many-worlds interpretation seem very horrible for most physicists. But any alternative is no less horrible. We are left with choosing between the lesser of two evils: discarding locality, or discarding realism. In fact, the Copenhagen interpretation refuses to interpret the main problem of the quantum formalism - the inexplicable "jump" from the state vector to the observable. Defending the Copenhagen interpretation Niels Bohr wrote: "*There is no quantum world. There is only an abstract quantum physical description. It is wrong to think that the task of physics is to find out how Nature is. Physics concerns what we can say about Nature*", the citation from[38]. In additional to this positivism point of view we can chosen only an idealism or realism one. An idealistic point of view, such as the information-theoretic interpretation, delivers from real non-locality. But according to this point of view physics investigates no reality but our (or no our?) knowledge about something. Such aim of physics was inadmissible for Einstein and it is inadmissible for many physicists, including the authors. Therefore we prefer to agree with Richard Feynman who wrote in the book[39]

"*There was a time when the newspapers said that only twelve mans understood of the theory of relativity. I do not believe that there was ever was such a time. ... On the other hand, I think it is safe to say that no one understand quantum mechanics. Do not keep saying to yourself, if you can possible avoid it, "But how can it be like that?" because you will get 'down drain' into a blind alley from which nobody has yet escaped. Nobody known how it can be like that.*"

But how can anybody try to make quantum computer if nobody understand quantum mechanics, i.e. if quantum formalism is not yet based on a generally accepted interpretation? It should note that most physicists do not think quantum mechanics needs interpretation, other than instrumentalist interpretations. The Copenhagen interpretation is the most popular among scientists, followed by the many worlds interpretations. But it is also true that most physicists consider non-instrumental questions (in particular ontological questions) to be irrelevant to physics. They fall back on Paul Dirac's expression: "shut up and calculate". It is interesting to note that Erwin Schrodinger, who shared with Dirac the Nobel Prize for 1933, stood up for the contrary opinion. He forewarned in the book[40] about a danger of "specialization barbarism". Unfortunately his apprehension has come true. The interpretation of the quantum parallelism, that is the main base of the idea the quantum computation, by many physicists[8] should be characterized by the Schrodinger term "specialization barbarism". It is no mere chance that David Deutch, the author of the idea of the quantum parallelism, argues forcefully against instrumentalism in this book[41]. We do not insist on the acceptability of the many worlds interpretations of the David Deutch[41] but everyone, who deals with the problem of the quantum computer, should realize the problem of the interpretation of quantum formalism in order to understand in which cases the quantum parallelism can not be possible.

### 3. WHY THE STERN-GERLACH EXPERIMENT CAN NOT BE CONSIDERED AS EXPERIMENTAL EVIDENCE OF QUANTUM BIT

The division on the "indifferent majority" and the "bothered minority" was suggested by N.D. Mermin in 1985[7], when only few physicists were versed in the problem connected with the Bell's inequality, see[42]. Now the Bell's paper is very popular. But the popularity can not guarantee understanding. Therefore the division suggested by N.D. Mermin remains almost invariable. Many physicists appertain to the "indifferent majority" heard only about the Bell's works and apparently therefore they are sure[8] that Bell has proved the impossibility of hidden-variables interpretations. Therefore it is needed to remind that only few physicists doubted in 1964 that quantum mechanics does not permit a hidden-variables interpretation because of the von Neumann's theorem published in 1932 in the book[23]. Bell, first of all, reconsidered[43] the von Neumann's no-hidden-variables theorem and had shown that it is based on unreasonable assumption. Moreover later, in 1988, he said in the Interview in Omni (see[10]): "Yet the von Neumann proof, if you actually come to grips with it, falls apart in your hands! There is nothing to it. It's not just flawed, it's *silly*! … When you translate [this assumption] into terms of physical disposition, they're nonsense. You may quote me on that: The proof of von Neumann is not merely false but *foolish*!"

In order to understand why the von Neumann's proof is silly it is needed to distinguish measurability and real existence. In fact, the main importance of the Bell's works[2,43] lies in the fact that they have clearly revealed this distinction. According to the Heisenberg's uncertainty principle observables with non-commuting operators can not be measure simultaneously and exactly. But why they can not be measure? The advocate of the Copenhagen interpretation assumed, at least in the first years, that it is impossible because of unavoidable disturbance of atomic system at any measurement. For example Paul Dirac wrote: "An act of observation is thus necessarily accompanied by some disturbance of the object observed… If a system is small, we cannot observe it without producing a serious disturbance and, hence we cannot expect to find any causal connection between the results of our observation". Niels Bohr repulsed Einstein's attacks on complementarity using the arguments of the unavoidable disturbance at measurement[29]. This notion about disturbance at measurement predominates among the "indifferent majority" up to now. But this disturbance presupposes just hidden variables existing before measurement. These variables are hidden since they can not be measured because of disturbance. In contrast to this the statement on realism violation based of the Bell test experiments[3-6] means that the outcome of a measurement is brought into being by the act of measurement itself and no preexisting value can be before the measurement.

3.1 Quantum parallelism and realism

Einstein stated that the positive philosophy of the Copenhagen interpretation results to solipsism. But in the reality it has resulted to scholasticism. Most physicists having studied the Copenhagen interpretation in the school (higher school) remain naive realists. Only few physicists can take seriously the questions: "Is the moon there when nobody looks?"[7] or even "Is the flux there when nobody looks?"[44] which bear a direct relation to the problem of quantum computation. Nobody proposed to use the moon for quantum computation for the present, but the superconductor loop considered by A. J. Leggett and A. Garg[44] was proposed in many papers[45-49]. It is important to understand that the affirmative (and natural for most people) answer on this question means impossibility of the quantum parallelism, i.e. the fundamental advantage of the quantum computation. The quantum parallelism can be possible since, according to quantum formalism and against of conventional logic, a system of $N$ elements (qubits), each of they is described by a single variable, can be describe by no $N$ but $2^N-1$ independent variables. It is possible only if each element of the system can not have any preexisting value before measurement. For example, if each spin-1/2 particle in a system described by the state vector

$$\psi = \gamma_1 \psi_\uparrow(r_1)\psi_\uparrow(r_2)... \psi_\uparrow(r_{N-1})\psi_\uparrow(r_N) + \gamma_2 \psi_\uparrow(r_1)\psi_\uparrow(r_2)... \psi_\uparrow(r_1)\psi_\downarrow(r_2)... + \gamma_n \psi_\downarrow(r_1)\psi_\downarrow(r_2)... \psi_\downarrow(r_{N-1})\psi_\downarrow(r_N) \quad (6)$$

has a preexisting spin projection $1/2$ or $-1/2$ with a probability $|\alpha_i|^2$ and $|\beta_i|^2$ correspondingly then the condition for the total probability $|\alpha_i|^2 + |\beta_i|^2 = 1$ should be applied to each $i$ particle and therefore only $N$ but no $n-1 = 2^N-1$ independent variables will be in the state vector (6). The condition $|\alpha_i|^2 + |\beta_i|^2 = 1$ should be applied even if the preexisting spin projection can not be measured, i.e. if it is the hidden variable. The same is valid also for the flux in superconductor loop considered in[44]. The quantum parallelism is not possible and any superconductor loop should not be proposed as a possible quantum bit if the flux can have a preexisting value when nobody looks, i.e. if the flux is the hidden variable. Thus, in order to understand how a quantum computer can not be made one must realize why the von Neumann's proof is silly and must distinguish measurability and realism, revealed in the term hidden variable. The variable is hidden if we can not measure it. But if a variable is not measurable it does not mean that it can not be real. And if a variable is real then the sum of probabilities of all permitted states must be equal unity $|\alpha_i|^2 + |\beta_i|^2 = 1$.

The title "Quantum Mechanics versus Macroscopic Realism: Is the Flux There when Nobody Looks?" of the paper[44] considering the problem of quantum superposition of macroscopically distinct states expresses clearly and correctly the contradiction between assumption of state superposition and realism presupposing variables (if even hidden) preexisting independently of measurement or observation. The EPR paradox has revealed the contradiction of superposition with local realism. Therefore in order to make a quantum computer with quantum parallelism (but no analog computer on base of a quantum system without the quantum parallelism) it is needed to break a principle of realism or locality. A quantum system can be considered as a possible quantum bit if it can not be described by a local hidden-variable theory.

3.2 Quantum superposition and realism

But many authors state that any two-state quantum system is quantum bit[45] and that already the famous Stern-Gerlach experiment gave evidence qubit existence in Nature[8]. In the Section III "Von Neumann's silly assumption" of the paper[10] David Mermin writes: "A third of a century passed before John Bell, 1966, rediscovered the fact that von Neumann's no-hidden-variables proof was based on an assumption that can only be described as silly - so silly, in fact, that one is led to wonder whether the proof was ever studied by either students or those who appealed to it". And now one should be led to wonder whether the Bell's paper was ever read by the authors stating that any two-state quantum system is qubit. Bell had shown that the von Neumann's proof is wanting on the example just of a two-state quantum system, a particle with spin-1/2.

The von Neumann's no-hidden-variables proof was based on the condition $<\Psi|s_x+s_z|\Psi> = <\Psi|s_x|\Psi> + <\Psi|s_z|\Psi>$. But there are absolutely no ground for imposing the requirement $<\Psi|s_x+s_z|\Psi> = <\Psi|s_x|\Psi> + <\Psi|s_z|\Psi>$ when the operators $s_x$ and $s_z$ do not have simultaneous eigenstate $\Psi$ and these properties cannot be simultaneously measured. It is incorrect to conclude that spin projections $s_x$ and $s_z$ can not really exist if they cannot be simultaneously measured. Their real existence does not contradict directly to quantum formalism predictions and all experimental results for a two-state system. Bell proposed in the paper[43] a simple example of a hidden-variable theory for a particle with spin-1/2 reproduces all predictions of superposition assumption about measurement result. This example means that no experimental result obtained on two-state quantum system can prove the existence of superposition and violation of realism.

### 3.3 Hidden-variables theory must be non-local and contextual

Having disproved the von Neumann's no-hidden-variables theorem John Bell has proved two others. According to the first one[2] in order do not conflict with the quantum formalism any hidden-variables theory must be non-local. According to the second one[43] any hidden-variables theory must be contextual. There is important to note that the experimental evidence[3-6] of violation of the Bell's inequalities should not be interpreted as a proof of impossibility of any hidden-variables interpretation in any case. Strictly speaking the Bell's experiments[3-6] can only proved that any hidden-variables interpretation must be non-local.

### 3.4 Entanglement and the Bohm's quantum potential

Bell's favorite example of a hidden-variables theory, the theory proposed by David Bohm[30], is not only explicitly contextual but explicitly and spectacularly non-local. In Bohm theory, which defies all the impossibility proofs, the hidden variables are simply the real configuration-space coordinates of real particles, guided in their motion by the wave function, which is viewed as a real field in configuration space. By using the conventional expression of the wave function $\Psi(r) = A\exp(i\varphi) = A\exp(iS/\hbar)$ Bohm[30] has divided the Schridinger equation

$$i\hbar \frac{\partial \Psi}{\partial t} = -\frac{\hbar^2}{2m}\nabla^2 \Psi + V\Psi \qquad (7)$$

into two, the almost classical Hamilton-Jacobi equation

$$\frac{\partial S}{\partial t} + \frac{(\nabla S)^2}{2m} + V - \frac{\hbar^2}{4m}\left(\frac{\nabla^2 P}{P} - \frac{(\nabla P)^2}{P^2}\right) = 0 \qquad (8)$$

and the conservation equation for the probability density $P = |\Psi|^2 = |A|^2$

$$\frac{\partial P}{\partial t} + \nabla\left(P\frac{\nabla S}{m}\right) = 0 \qquad (9)$$

where $\nabla S = p$ is a momentum and $\nabla S/m = v$ is a velocity. The equation (8) differs from the classical Hamilton-Jacobi one because of the term $Q_B = -(\hbar^2/2m)(\nabla^2 P/P - (\nabla P)^2/P^2)$ called by Bohm "quantum potential"[30]. In contrast to the conventional potential $V$ the Bohm's quantum potential $Q_B$ is hideously non-local, as well as the entanglement (the EPR correlation). The quantum potential $Q_B$ can be used for the description of the entanglement and understanding of hideous paradoxicality of this base of the quantum computation.

Verner Heisenberg stated[50] that the Bohm's interpretation[30] does not differ from the Copenhagen interpretation according to positivism point of view. It is correct. But because of the invincible realism of most physicists the Bohm's interpretation is more convenient in order to understand, what is entanglement. The non-local quantum potential it is very strange, but it is more intelligible for any realist than entanglement of our knowledge.

### 3.5 The Bohm quantum potential and the Aharonov-Bohm effect

The non-local quantum potential can describe quantum non-locality including the non-locality becoming apparent in the double-slit interference experiments, see Section 2. The Bohm's non-local hidden variable theory reinterprets this experiment, and others like it, in terms of particles moving along perfectly definite trajectories. The only difference is that these trajectories are influenced, not only by the slits themselves, but also by the quantum potential[51]. Y. Aharonov and D. Bohm have shown in 1959[52] that electromagnetic potentials can change the interference pattern. The Aharonov-Bohm effect arises directly from the quantum formalism. The phase difference $\varphi_1 - \varphi_2$ in the relation (2) for the interference of a particle with a charge $q$ should depend both from the electrostatic potential $V$, $\varphi_1 - \varphi_2 = \Delta\varphi_0 + qVt/\hbar$ and from the magnetic vector potential $A$ (from the magnetic flux $\Phi$) $\varphi_1 - \varphi_2 = \Delta\varphi_0 + q\Phi/\hbar$[29] since $\partial\varphi/\partial t = E/\hbar$, $\nabla\varphi = p/\hbar = (mv+qA)/\hbar$ and the energy of the particle $E = (p-qA)^2/2m + qV$. But it is no mere chance that the paper[52] predicting this effect and the paper[30] introducing the quantum potential were written the same author. The Aharonov-Bohm experiments demonstrate that the interference pattern can be altered without any force having acted on the particle, if the quantum potential is not take into account.

But only few physicists can accept a real non-locality and a real non-local quantum force -$\nabla Q_B$ acting on the particles. In order to explain the non-local force-free quantum momentum transfer observed in the Aharonov-Bohm experiments some authors[53] propose to reconsider the principle of complementarity.

## 4. HAS IT BEEN PROVED THAT A SUPERCONDUCTING LOOP COULD BE FLUX QUBIT?

Since the quantum parallelism is based on the EPR correlation (the entanglement) contradicting to the local realism in order to make quantum computer a principle of realism or locality should be violated. The experiments with photons[3-6] and other prove the invalidity of realistic local theories only on the level of elementary particles for the present. But there is few chance that technology will be able to work on this level in the near future. Therefore the only chance to make the quantum computer in the near future should be connected with the level higher than atomic one, the level is mastered already by the modern technology. Superconductivity is attractive for a realization of the idea of quantum computer since it is one of the macroscopic quantum phenomena. In order to make a quantum computer on base of superconductor structures it is needed to understand how entangled states can be created in superconductor structure and to prove experimentally a possibility of quantum superposition of macroscopically distinct states.

### 4.1 Proposals by authors appertain to the "indifferent majority" to create entangled states with help of a classical interaction

The numerous proposals to couple qubits with help of a classical interaction[45,54,55] and the statements on experimental evidence of quantum gate[56] and entangled states[57,58] are most impressing consequence of the "specialization barbarism". The paradoxical nature of the entanglement is exposed in the EPR paradox. The long history of this paradox demonstrates clearly that if we have studied anything it does not mean that we understand this. It should be obvious for any physicists that no classical interaction can be used for a creation of a quantum register since the EPR correlation (the entanglement) is so paradoxical phenomenon that we must choose between a real non-locality[30] and the absence of an objective reality. No classical interaction can force to make such paradoxical choice. But some physicists appertain to the "indifferent majority" ignore the paradoxical nature of the EPR correlation and are sure that it is enough to write a Schrodinger equation (7) in order to obtain the entanglement. There is important to emphasize that no interaction in the Schrodinger equation except for the non-local Bohm's quantum potential in (8) can create the non-local EPR correlation since any other interactions are local in essence. It is not clear now how the Bohm's quantum potential can be created in superconductor structures. Both in the proposals[45,54,55] and in the experiments[56-58] the quantum potential is absent.

### 4.2 Bell's example of hidden-variables interpretation and "experimental evidence" of quantum superposition of macroscopically distinct states

The authors[44] indicating the contradiction of an assumption on quantum superposition of macroscopically distinct states with macroscopic realism and noninvasive measurability at the macroscopic level propose inequalities similar to those of Bell[2] or of Clauser et al.[59] for experimental test of this contradiction. But L. E. Ballentine[60] has argued that there can not be revealed contradiction of quantum formalism with realism in the two-state system considered in[44], and the contradiction is only with noninvasive measurability. Indeed, Bell had shown[43] that no experiment can reveal a contradiction between realistic and quantum formalism prediction in the case of a single spin-1/2, when different spin projections can be measured. In contrast to spin-1/2 an SQUID, i.e. superconductor loop interrupted with Josephson junction, considered in[44] has only projection. Therefore all experimental results obtained on this two-state quantum system can be easy interpreted in the term of local hidden variables. A.J. Leggett would like to think[61] that the experimental results[62,63] can be interpreted as an experimental evidence of superposition of macroscopically distinct quantum states. But this wish contradicts to the Bell's example[43] of hidden-variable interpretation according to which none of the experimental results[62-69] made on any two-states quantum system can give evidence of quantum superposition violating the principle of realism. Moreover the authors[70] state that the Rabi-oscillations, the Ramsey-fringes and other effects interpreted as experimental evidence of quantum superposition in superconductor loop can be interpreted classically. Thus no experimental evidence of quantum bit in superconductor structures was obtained for the present and it is funny that the company D-Wave Systems "demonstrated" already the world's first quantum computer made on base of superconductor loops.

### 4.3 Hidden variables can become no-hidden

Both quantum superposition and the hidden-variable interpretation[43] presuppose that a single-shot measurement of a two-state system should give a result corresponding one of the two permitted values and a multiple measurement should give an average value. For example the single-shot measurement on spin projection along a direction **n** of the eigenstate $\psi = \psi_\uparrow$ of a spin-1/2 particle along **z** should give *1/2* or *–1/2* and the multiple measurement should give *1/2cos$\phi$*, where $\phi$ is the angle between **n** and **z**. These results should be observed according to both the superposition and hidden-variable interpretation[43]. But according to the superposition interpretation no projection value exists before the measurement whereas according to the hidden-variable interpretation[43] spin projection has a random value which changes to one of the permitted value *1/2* or *–1/2* at the measurement.

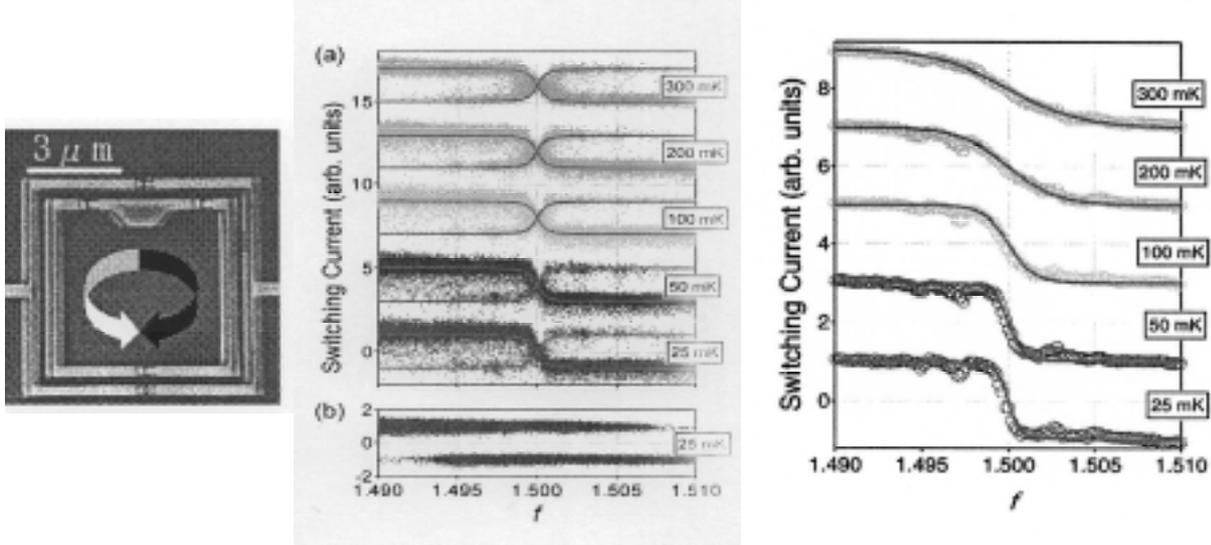

Fig. 2. The magnetic dependencies on the external magnetic flux $f = \Phi_{ext}/\Phi_0$ of the switching current of the superconductor interferometer (the superconductor loop interrupted with two Josephson junctions, see the right figure) corresponding to the additional magnetic flux $\Delta\Phi_{lp} = LI_p$ induced by the persistent current $I_p$ of the loop interrupted with three Josephson junctions located inside the interferometer, see the right figure, at different temperatures. The points shown on the central figure correspond to the single-shot measurements. At the left the results of the single-shot measurements taken an average at each magnetic flux $f$ value. The solid curves represent thermal-averaged theoretical value.

The superconductor loop with the magnetic flux $\Phi$ equal half of the flux quantum $\Phi_0 = \pi\hbar/e$ is considered as flux qubit and an analog of a spin-1/2 particle[45]. The loop interrupted with three Josephson junctions is investigated in the most papers[71-76] as the flux qubit. Because of the demand of the current conservation $I_p = I_{c1}sin(\Delta\varphi_1) = I_{c2}sin(\Delta\varphi_2) = I_{c3}sin(\Delta\varphi_3)$ and the relation $\Delta\varphi_1 + \Delta\varphi_2 + \Delta\varphi_3 + 2\pi\Phi/\Phi_0 = 2\pi n$ between the phase difference $\Delta\varphi_1, \Delta\varphi_2, \Delta\varphi_3$ between boundaries of each Josephson junctions and the magnetic flux $\Phi$ inside the loop a stable state with zero persistent current $I_p = 0$ is forbidden at $\Phi = (n+0.5)\Phi_0$ and there is two permitted states *n* and *n+1* with the persistent current having the same value and opposite direction $I_p(n+1) = -I_p(n)$. The multiple measurements of the additional flux $\Delta\Phi_{lp} = LI_p$ induced by the persistent current $I_p$ in the loop interrupted with three Josephson junctions having non-zero critical current $I_{c1} \neq 0$, $I_{c2} \neq 0$, $I_{c3} \neq 0$, give the result corresponding to the expected average value $\Delta\Phi_{lp} = L<I_p> = L[I_p(n+1) + I_p(n)]/2 = 0$ at $\Phi = (n+0.5)\Phi_0$ in all works[71-74]. But the single-shot measurements give the expected results $\Delta\Phi_{lp} = LI_p(n) \neq 0$ and $\Delta\Phi_{lp} = LI_p(n+1) \neq 0$ only in some works[75]. In other work[76] the magnetic dependencies $LI_p(n)$ and $LI_p(n+1)$ on $f = \Phi_{ext}/\Phi_0$ (the external flux $\Phi_{ext} >> \Delta\Phi_{lp}$) show a $\chi$-shaped crossing at $\Phi_{ext} = 1.5\Phi_0$, Fig.2. The authors[76] interpret this $\chi$-shaped crossing as a quantum behavior and its absence as the classical behavior. But this interpretation can not be correct. The value $\Delta\Phi_{lp} = LI_p = 0$ is forbidden at $\Phi_{ext} = 1.5\Phi_0$ and no forbidden value can be observed according to both the superposition and hidden-variable interpretation. The $\chi$-shaped crossing $\Phi_{ext} = 1.5\Phi_0$, Fig.2, is no quantum behavior but it is a paradoxical behavior contradicting to the quantum formalism. This experimental challenge to the quantum formalism at the macroscopic level can not eliminated by the assumption that the results correspond no the single-shot measurements but an average value at $\Phi_{ext} = 1.5\Phi_0$. The width of the $\chi$-shaped crossing does not depend on temperature and much smaller the width of the change from 0 to 1 of the *n* state probability increasing with temperature, Fig.2, because of thermal fluctuation.

4.4 The two states predicted by quantum formalism in superconductor loop are not observed in some cases

Other experimental challenge to the quantum formalism was discovered at measurements of the quantum oscillation in the magnetic field of the critical current of asymmetric superconductor loops[77]. The critical current should change periodically in magnetic field $\Phi$ with the period corresponding to the flux quantum $\Phi_0$ inside the loop since the external measuring current $I_{ext}$ is added in one of the loop halves with the circular persistent current $I_p = 2I_{p,A}(n - \Phi/\Phi_0)$, the direction and value of which are periodical function of $\Phi/\Phi_0$, Fig.3. The critical current, $I_{c+}$, $I_{c-}$, measured in opposite direction, should be equal

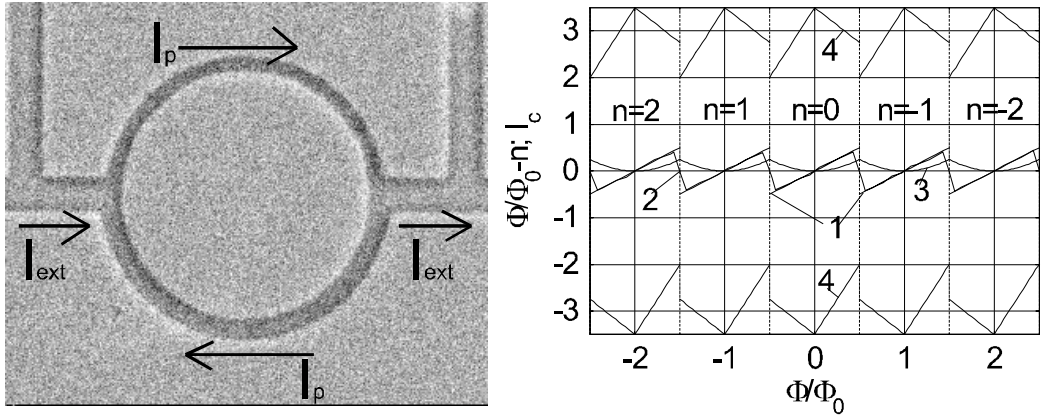

Fig. 3. At the right: SEM image of one of the asymmetric Al rings with radius $r = 2$ $\mu m$ used in the work[77]. The measuring $I_{ext}$ and the persistent $I_p$ currents are added in the wide or narrow half depending on their directions. At the left: Magnetic dependencies corresponding to the lowest permitted state of the persistent current $I_p = 2I_{p,A}(n-\Phi/\Phi_0)$ (1), the average equilibrium $I_p$ value $<I_p> = 2I_{p,A}(<n> - \Phi/\Phi_0)$ (2), the square of the persistent current $I_p^2$ (3), and the critical current, $I_{c+}$, $I_{c-}$, expected according to the relations (10) for a loop with $s_w/s_n = 2$, $I_{c0} = 3.5$ $\mu A$ and $I_{c0} = 3.5$ $\mu A$.

$$I_{c+}, I_{c-} = I_{c0} - 2I_{p,A} | n - \frac{\Phi}{\Phi_0} | (1 + \frac{s_w}{s_n}) \tag{10a}$$

if $I_{ext}$ and $I_p$ are added in the narrow half, Fig.3, with the section $s_n$ and

$$I_{c+}, I_{c-} = I_{c0} - 2I_{p,A} | n - \frac{\Phi}{\Phi_0} | (1 + \frac{s_n}{s_w}) \tag{10b}$$

if $I_{ext}$ and $I_p$ are added in the wide half with the section $s_w$. According to the relation (10) the measurement of the critical current of asymmetric loop, corresponding to the single-shot measurement, should reveal two states *n* and *n+1* at $\Phi = (n+0.5)\Phi_0$, Fig.3. But the experiment[77] has shown that only the multiple measurements of the values proportional $<I_p>$ and $I_p^2$ corroborate the two states *n* and *n+1*, whereas measurements of the critical current oscillations, $I_{c+}(\Phi/\Phi_0)$, $I_{c-}(\Phi/\Phi_0)$, have given very strange result contradicting to the quantum prediction of the two states *n* and *n+1* at $\Phi = (n+0.5)\Phi_0$. The maximums and minimums of the oscillations $I_{c+}(\Phi/\Phi_0)$, $I_{c-}(\Phi/\Phi_0)$ are observed no at $\Phi = n\Phi_0$ and $\Phi = (n+0.5)\Phi_0$ as should be according to the quantum formalism, Fig.3 but at $\Phi = (n+0.25)\Phi_0$ and $\Phi = (n+0.75)\Phi_0$.[77]

## 5. CONCLUSION

The company D-Wave Systems promises to make a quantum computer for commercial applications in the nearest future on base of superconductor nano-structures. Indeed, superconductivity is most real way to achieve success. But unfortunately research workers of the D-Wave Systems seem to ignore the fundamental difference between the quantum computer and the analog computer made on base of any quantum system. They, as well as many physicists, refuse to understand that in order to make the effective quantum computer a realism principle should be violated. In order to make the quantum computer on base any macroscopic quantum system the principle of macroscopic realism should be violated. Therefore the main task is now the search of an experimental evidence of this violation. The results obtained on the atomic level may be useless in this search since the experiments testify fundamental differences between application of basic principles of quantum mechanics on atomic and macroscopic levels[78].

## ACKNOWLEDGMENTS

This work was financially supported by a grant "Quantum bit on base of micro and nano-structures with metal conductivity" in the Program "Technology Basis of New Computing Methods" of ITCS department of Russian Academy of Sciences and by the Presidium of Russian Academy of Sciences in the Program "Quantum Nanostructures".